\documentclass[
preprint,
 amsmath,amssymb,
 aip
]{revtex4-1}

\usepackage{graphicx}
\usepackage{dcolumn}
\usepackage{bm}
\usepackage[font=small,labelfont=bf]{caption}
\usepackage{enumerate}

\long\def\@makecaption#1#2{
  \vskip\abovecaptionskip
  \sbox\@tempboxa{#1: #2}
  \ifdim \wd\@tempboxa >\hsize
    #1: #2\par
  \else
    \global \@minipagefalse
    \hb@xt@\hsize{\hfil\box\@tempboxa\hfil}
  \fi
  \vskip\belowcaptionskip}

\begin{document}

\title{Bipolar electrical switching in metal-metal contacts\\}

\author{Gaurav Gandhi}
 \email{gaurav@mlabs.in}
\author{Varun Aggarwal}%
\email{varun@mlabs.in}

\affiliation{
 mLabs, New Delhi, India
}

\begin{abstract}
Electrical switching has been observed in carefully designed metal-insulator-metal devices built at small geometries. These devices are also commonly known as memristors and consist of specific materials such as transition metal oxides, chalcogenides, perovskites, oxides with valence defects, or a combination of an inert and an electrochemically active electrode. No simple physical device has been reported to exhibit electrical switching. We have discovered that a simple point-contact or a granular arrangement formed of metal pieces exhibits bipolar switching. These devices, referred to as coherers, were considered as one-way electrical fuses. We have identified the state variable governing the resistance state and can program the device to switch between multiple stable resistance states. Our observations render previously postulated thermal mechanisms for their resistance-change as inadequate. These devices constitute the missing canonical physical implementations for memristor, often referred as the fourth passive element. Apart from the theoretical advance in understanding metallic contacts, the current discovery provides a simple memristor to physicists and engineers for widespread experimentation, hitherto impossible.
\end{abstract}

\maketitle

\section{\label{sec:level1} INTRODUCTION}
Leon Chua defines a memristor as any two-terminal electronic device that is devoid of an internal power-source and is capable of switching between two resistance states upon application of an appropriate voltage or current signal that can be sensed by applying a relatively much smaller sensing signal \cite{chua2011resistance}. A pinched hysteresis loop in the voltage vs. current characteristics of the device serves as the fingerprint for memristors. Despite the simplicity of symmetry argument that predicts the existence of memristor \cite{strukov2008missing, chua2011resistance}, no simple physical device behaving as a memristor has been observed so far and thus considered to be non-existent \cite{pershin2011teaching}. Current memristor implementations use specialized materials such as transition metal oxides, chalcogenides, perovskites, oxides with valence defects, or a combination of an inert and an electrochemically active electrode.

\begin{figure}[h]
\centering
\includegraphics[angle=-90, scale=0.35]{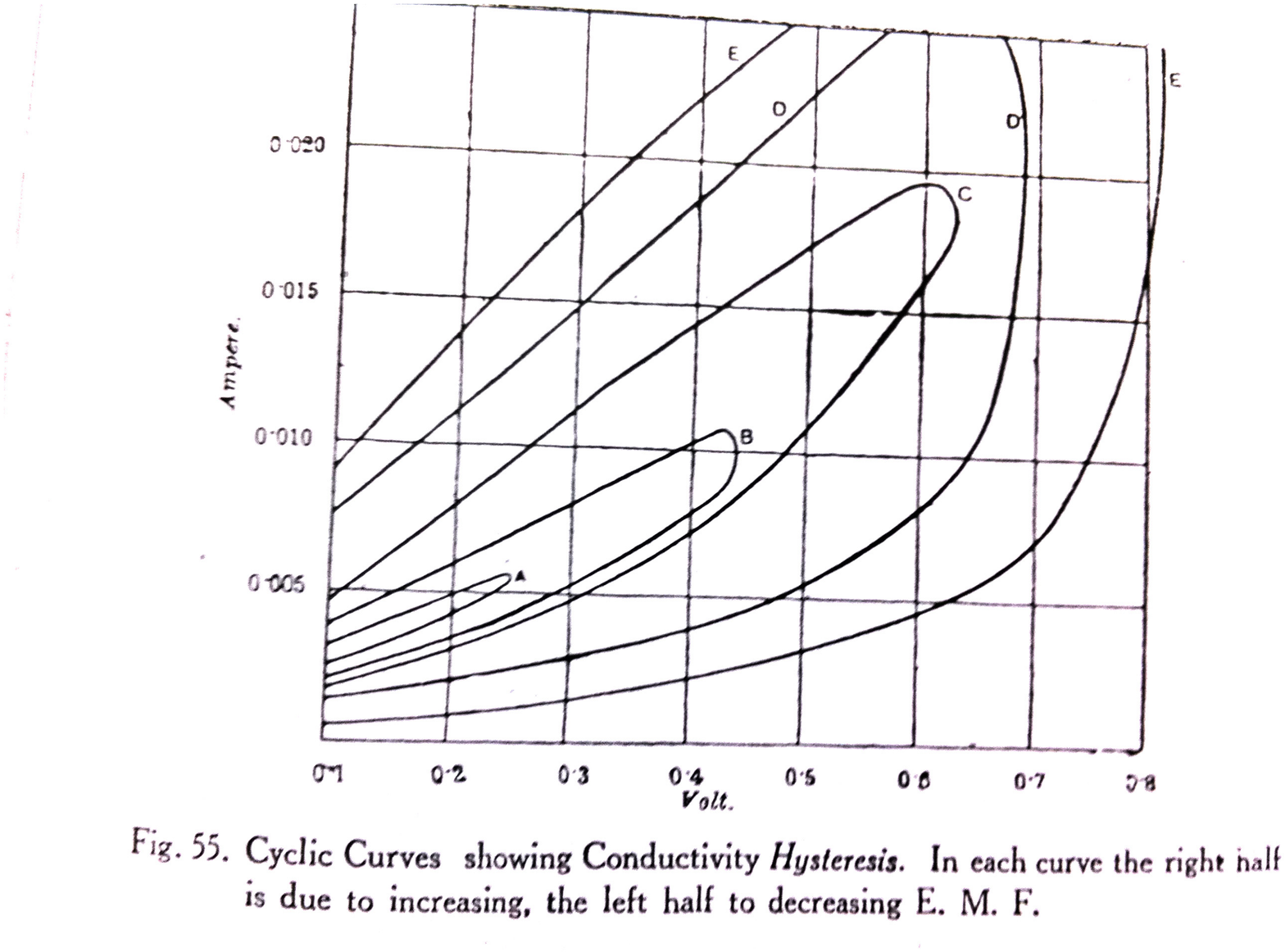} 
\caption{Bose's observation of pinched hysteresis in current (vertical axis) and voltage (horizontal axis) for iron filing coherer. Interestingly, this reference has been missed by almost all the papers on memristors.} 
\label{fig_bose} 
\end{figure}

On the other hand, coherer, invented by Edouard Branly \cite{dilhac2009edouard,lodge1897history, falcon2010efecto} in the 19th century, in its many embodiments such as ball bearings, metallic filings (also referred to as granular media) in a tube or a point-contact exhibits an initial high-resistance state and coheres to a low-resistance state on the arrival of radio waves. The device attains its original resistance state on being tapped mechanically. The first electrically reset-able coherer, comprising a metal-mercury interface and named as an auto-coherer, \cite{bose1901change,bose1904detector, bose1899self, bondyopadhyay1998sir} did not require tapping and resets in the absence of radio waves. Although Bose observed that coherers demonstrate a pinched hysteresis I-V curve in the first quadrant (arguably the first such observation; Ref. \ref{fig_bose}) \cite{bose1901change} and exhibit multiple stable resistance-states, he could not establish a systematic way to electrically reverse the diminution of resistance \cite{bose1899electric} \footnote{Cat's whisker was the first metal-semiconductor point contact device patented by JC Bose and was actively used in early radio research}.

\begin{figure}[h]
\centering
\includegraphics[scale=0.45]{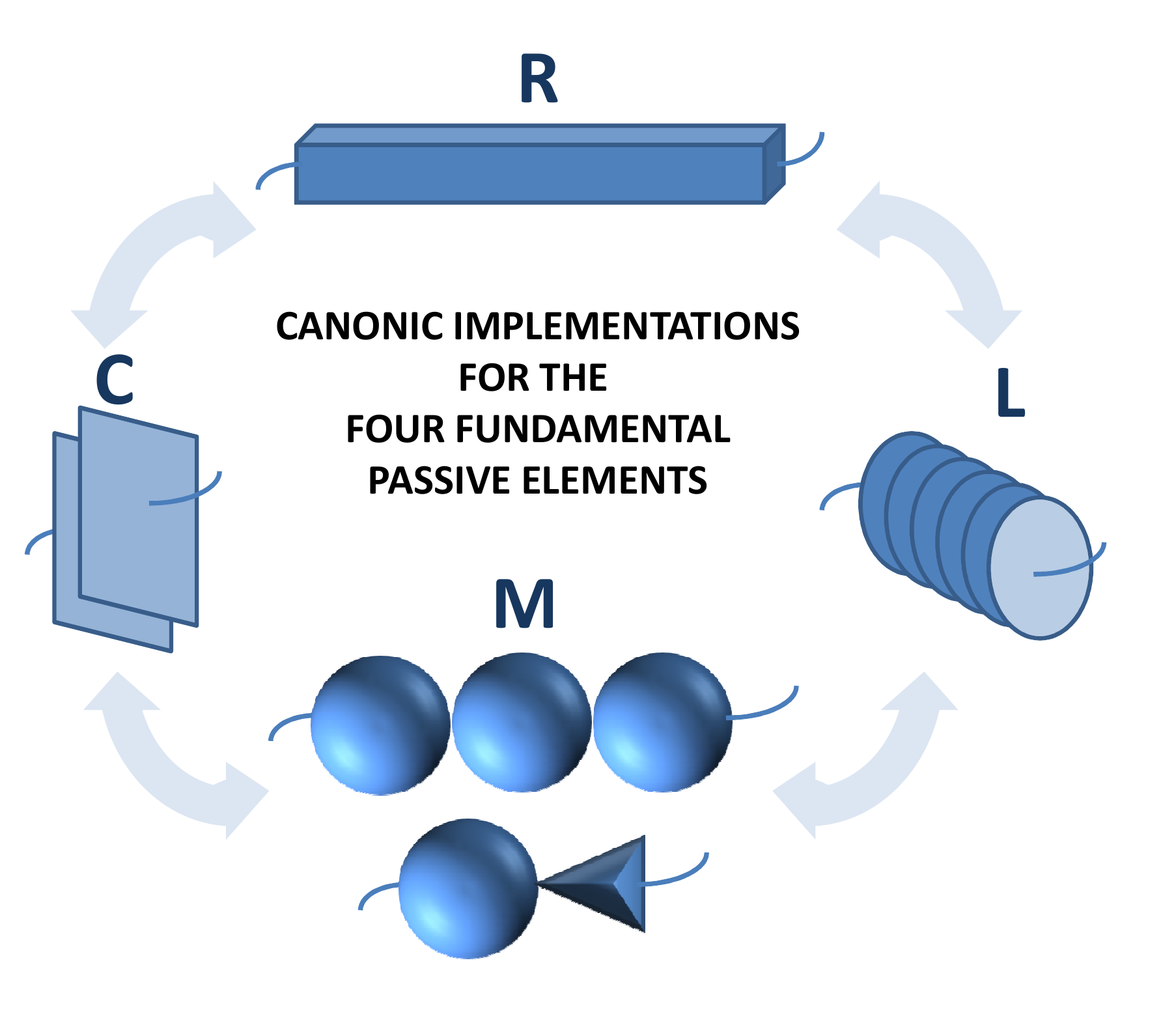} 
\caption{Complete set of canonic discrete implementations of the four fundamental circuit elements.} 
\label{canonicfig} 
\end{figure} 

Among several competing theories for explaining the coherer behavior, such as joule heating, molecular rearrangement, Seebeck and Peltier Effect \cite{bose1901change, ecclesthesis, eccles1912electrothermal, lodge1901oliver, bquin2010electrical}, the most popular theory was that of current-induced heating resulting in the welding of metal-metal contacts that led to diminution of resistance. For the auto-coherer, Eccles \cite{eccles1909coherers} postulated that current leads to the heating of oxide at the interface contacts and the change in resistance is a function of the temperature of the oxide. His thermistor equation for the said behavior is, in principle, the same as that prescribed by Chua and Kang for a thermistor \cite{chua1976memristive}, and satisfies the conditions of Chua's memristive one-port \cite{chua1971memristor}. The equation proposed by Eccles is being reproduced here:
 \begin{eqnarray}
\frac{d\theta}{dt}=k\rho c^{2}-m \theta \\
L \frac {dc}{dt} +(r+\rho)c = \epsilon \\
\rho = \rho_{0}(1+\alpha \theta) 
\end{eqnarray}

Where c represents the current, $\rho$ represents the resistance of oxide, $\epsilon$ represents the voltage, $\theta$ is the temperature of the oxide  and other variables pertain to the setup mentioned in \cite{eccles1909coherers}.  L and r refer to the resistance and inductance in series with the coherer.

Thus the existence of electrically-controllable multiple resistance-states, and the possibility of a memristive constitutive relationship was known over a century ago. Though no one (including Eccles and Bose) observed a pinched hysteresis in both the quadrant.
In works related to coherer in the last decade, its bi-stability has been reported and its multi-stable behavior has been confirmed \cite{bquin2010electrical, falcon2005electrical}. A thermal mechanism, similar to numerous others proposed a century back, has been postulated to explain the resistance change. All these studies affirm the unidirectionality of the resistance value (which fatigues with time) and propose no method to electrically recover the older, higher resistance states of the device. On the other hand, autocoherer has been shown to exhibit diode-like rectifying properties \cite{philips1998italian, groenhaugexperiments}. 
In the present work we have established, by uncovering hitherto unknown electrical properties of a set of coherer and autocoherer, that extremely simple devices show memristive properties. We have found that the coherer and the auto-coherer are electrically-controllable state-dependent resistors, the state variable being the maximum current flown through the device. We have, for the first time observed bipolar switching in these devices, wherein the device can indeed be programmed (electrically) to an older higher resistance state. The state-map of the resistance of the device vs. current is different for the two directions of the current (Ref. Fig \ref{coherer}(d)), which allows to write and erase it as a memory. The programmed resistance of the device can be read by another characteristic signal of small amplitude. Analogous to the phenomena of a wire showing resistive properties, a coil being inductive, and a set of conducting plates separated by a dielectric exhibiting capacitance, we show that two convex metallic surfaces in contact are memristive in nature and work as a fully-functional resistive RAM (Refer Fig. \ref{fig_imax}).

It is worth discussing what causes the resistance switching. The cause could be existence of certain impurities or oxide at the interface or merely by the geometry of the interface. We note, however, that no Metal-Insulator-Metal system containing just oxide and iron, or those built in a macro-dimension, has been reported to show memristive properties. We analytically deduce that the switching in our devices cannot be due to the different mechanisms observed in oxide-based bipolar memristors. Based on our material setup and experimental observations, our best understanding is that the resistance switching is caused by electric-field induced polarization at the interface of the metals (which may or may not contain impurities). This is discussed in detail in a later section.
The paper makes the following contributions: 
\begin{enumerate}[(i)]
\item Completes the entire set of canonic implementations of all the four known passive elements of circuit theory, 
\item Reports for the first time bipolar switching in simple metallic constructions indicating the ubiquity of the memristive phenomena, 
\item Argue that thermal mechanism of resistance change in metallic contacts is inadequate and hypothesize a electric-field polarization as its cause,  
\item Shows that memristor phenomena is not limited to specific materials assembled at small geometries, but is present in a large class of metals put together as a point contact and 
\item Provide an inexpensive and simple memristor for widespread experimentation, hitherto impossible.
\end{enumerate}

This paper is organized as follows: Section II describes the constructions of three embodiments of the devices. These include those with a point contact between metals, a granular media assembly and a third comprising of a metal in liquid form. Section III describes in detail the electrical properties of these devices and their behavior under different electrical stimulations. Based on the observed behavior, we postulate an electrical model for the devices and identify the state-variable controlling the resistance change.  In Section IV, we discuss the implication of our observations, hypothesize the physical mechanism governing the behavior of the devices and compare it with other memristor devices.

\section{\label{sec:level1} MATERIALS AND METHODS}

\begin{figure}
\centering
\includegraphics[scale=0.55]{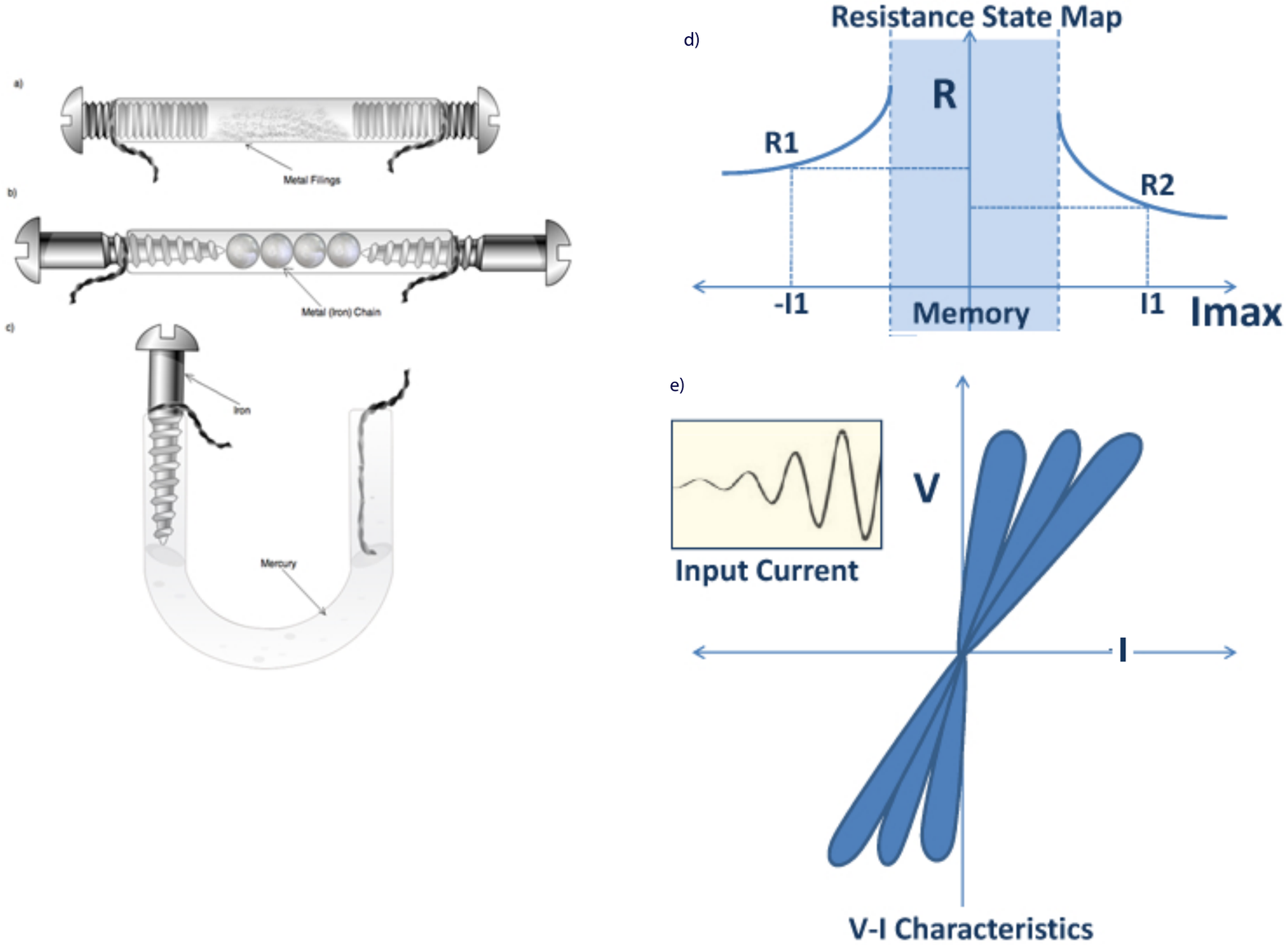} 
\caption{{\bf LEFT: Various embodiments of coherer used for experimentation. (a)} Iron Filing Coherer (IFC) {\bf (b)} Iron Chain Coherer (ICC) {\bf (c)} Iron Mercury Coherer (IMC).  {\bf RIGHT: (d)} Resistance State Map for the device. Here horizontal axis refers to maximum current that has flowed through the device while vertical axis is the resistance of the device.\footnote {It shows that the current devices exhibit a state-dependent, electrically-controllable resistance, the state variable being the maximum current that has flowed through the device. The state-map is different according to the direction of the current, which enables us to switch the devices to-and-fro between multiple resistance states. The resistance of the device can be read in the memory state without altering it.} {\bf (e)} Current-voltage characteristics of the device for a current-mode sine wave signal of increasing amplitude. The device shows the famous pinched hysteresis loops and various possible current-voltage values for the same current. } 
\label{coherer} 
\end{figure}

The current section discusses the construction of the devices which can be accomplished in any simple undergraduate electrical engineering lab. We replicated three embodiments of the coherer and autocoherer: an Iron Filing Coherer (IFC), an Iron Chain Coherer (ICC), and an Iron Mercury Coherer (IMC) (see Fig. \ref{coherer}(a-c) ) \footnote{The experiment was repeated with several metals, including aluminum and magnesium flakes and nickel and zinc-coated ball bearings.  Qualitatively similar results as reported herein were observed in all these experiments.}.

The first embodiment, namely, Iron Filing Coherer (IFC), consists of a tube containing closely-packed iron filings with electrodes in contact with the metal filings at the two ends of the tube. In the second embodiment, called Iron Chain Coherer (ICC), iron filings are replaced by a chain (linear assembly) of iron beads and the third embodiment is an embodiment of the self-recovering coherer consisting of a U-tube filled with mercury forming contact with an iron screw on one side. In the third embodiment, henceforth referred as Iron Mercury Coherer (IMC), one electrode is connected to an iron screw, whereas the other dips into mercury on the other side of the U-tube. Depending on the packing density (IFC), pressure applied (ICC) and contact area (IMC), the devices show a continuum of states between a nonlinear high-resistance state and a more linear low-resistance state.  The next section discusses the electrical behavior exhibited by the three devices.

\section{\label{sec:level1}  EXPERIMENTAL RESULTS }

These devices were activated by different current-mode input signals in their non-linear mode, and their transient behavior was recorded. We found that the three devices show similar qualitative behavior and that the mercury-iron system does not function as a diode, as previously reported \cite{philips1998italian, groenhaugexperiments}, but exhibits state-dependent resistance. All the observed behavior is common to the three devices. We have found that the devices exhibit three distinct behaviors: Cohering action, multi-stable memristive behavior, and bistable resistive RAM behavior. 

\subsection{\label{sec:level2}  Cohering Action} 

For any input current leading to a voltage below a specific threshold voltage, $V_{th}$, the devices exhibit a high non-linear resistance and may be used for rectification. Whereas IMC readily shows a moderate non-linear resistance that can be used for demodulation, IFC and ICC require considerable adjustment to do so. Due to this, only the IMC has historically been used for demodulation. In this region, the device remembers the resistance it had earlier, and continues to exhibit the same. We call this region as the memory state.  

At a current higher than $I_{th}$, corresponding to a voltage $V_{th}$, the resistance of the device falls sharply (Refer P1 transition in Fig. \ref{fig_imax}), and the device exhibits lower conductance. Once the device takes this new state, it maintains the said resistance on excitation by current values below $I_{th}$ as well. This is the well-known cohering behavior used for detecting electromagnetic waves. The device cannot be reset electrically to a resistance less than that shown at A2 (Ref. Fig. \ref{fig_imax}). Contrary to earlier observations, this behavior is also exhibited by IMC \cite{philips1998italian}. 

\subsection{\label{sec:level2}  Multistable Memristive behavior} 

Once cohered, the device exhibits a state-dependent resistance, the state variable being the maximum current ($I_{max}$), i.e. $R_{t} = f([I_{max}]_{0-t})$. As the device is exposed to pulses of subsequently larger peak current (Refer Fig. \ref{fig_imax}, \footnote {Note that the resistance changes appreciably only when the maximum current through the device has changed. This can be seen through color correspondence, where each color shows a new stable resistance-state and the resistance transitions are marked by the first pulse of higher amplitude: P1, P2, P3, P4 and P5. In case the maximum current passed through the device does not change, the resistance feebly oscillates around the same value, as seen in the region of A1, A2, A3, A4 and A5. Furthermore, one may observe that the resistance remains fixed even when the amplitude of the pulse is decreased (A6) , since the maximum current has not changed.}), it sets itself to new resistance values. The resistance remains non-linear, nonetheless. The maximum voltage across the device remains practically constant at $V_{th}$. This behavior is akin to that of a diode, but unlike a diode the device remembers its changed resistance when taken to lower voltage levels. For input current pulses of same or lower amplitude than the maximum current experienced, the device shows hysteresis loops around the already-achieved resistance value, with small oscillations. In \cite{bquin2010electrical}, some of these behaviors have been observed for ICC.  

\subsection{\label{sec:level2}  Bistable Resistive RAM} 

 We have found that the resistance of the device is a function of the magnitude of $I_{max}$ for either directions of current, but with a quantitatively different state-map, making it behave as a resistive RAM. This can be mathematically stated as: 

 \noindent Let 
 \begin{eqnarray} 
 \noindent R_{p1} = f(magnitude([I_{max+}]_{0-t})) = I_{1}), \\  
 R_{n1} = f(magnitude([I_{max-}]_{0-t}))=I_{1}), \\ 
\implies
 R_{p1} \neq R_{n1}
 \end{eqnarray} 
where $R_{p1}$ is the resistance of the device when activated by a maximum current of $I_1$ in positive direction, and $R_{n1}$ is the resistance when activated by a maximum current of $I_1$ in the negative direction. $f(magnitude([I_{max+}]_{0-t}))$ implies the maximum current the device has experienced between time=0 and time=t. (Ref. Fig. \ref{coherer})

\begin{figure}
\centering		
 \includegraphics[angle=-90, scale=0.7]{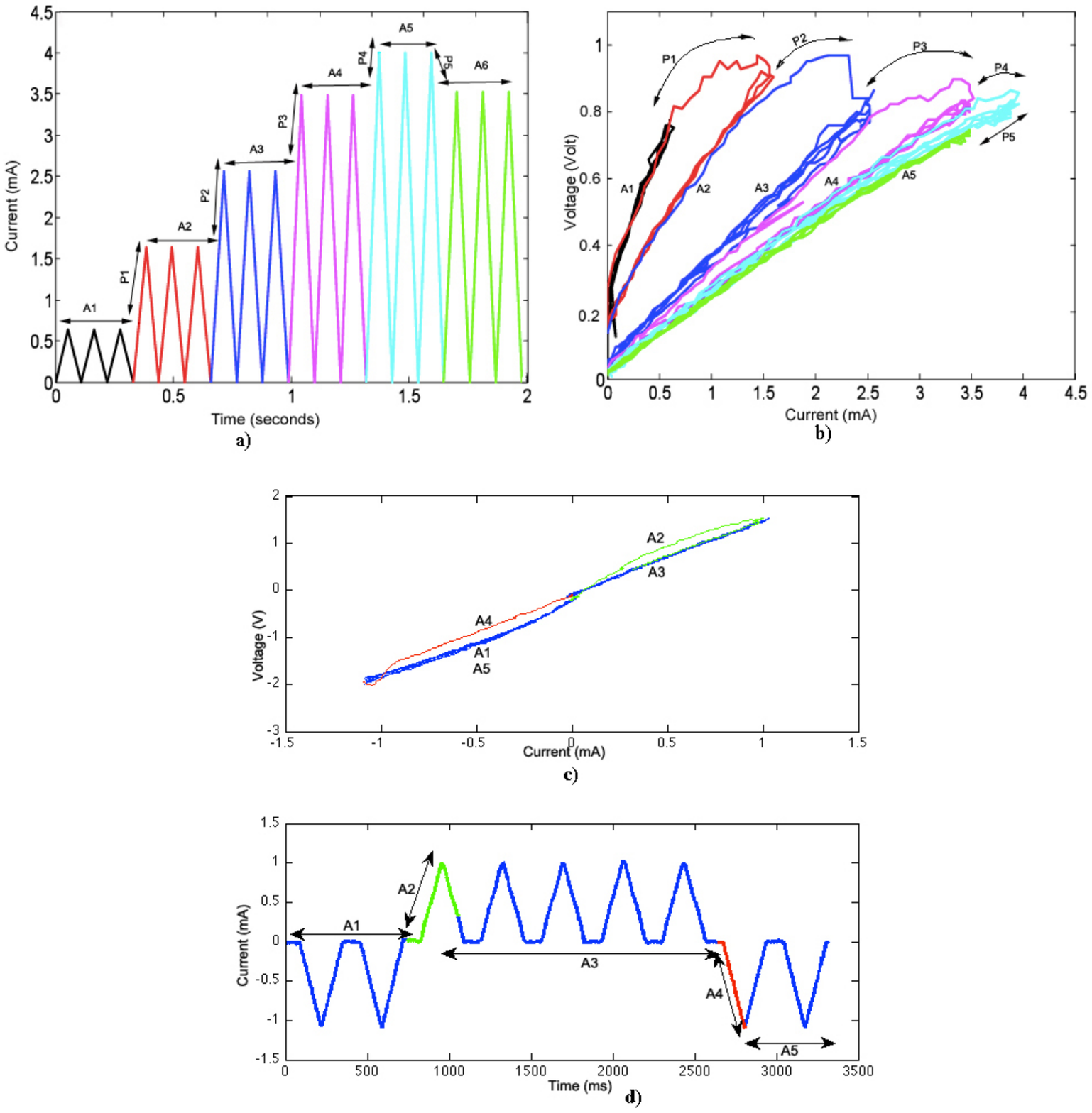} 
 \caption{{\bf Device behavior as a state-dependent resistance. a:} Input Current vs. Time and {\bf b:} Current-Voltage Plot. After configuring the device in the non-linear high-resistance mode, an input current pulse with varying amplitude is applied across it. It is observed that the maximum voltage across the device does not cross a threshold voltage, $V_{th}$.\\
 {\bf Bistable RRAM behavior} (c) Input current vs. time, (d) Voltage across device vs. input current. One clearly observes pinched hysteresis loop for Iron Filing Coherer. }
 \label{fig_imax} 
 \end{figure}

When activated by any two-sided current input, the device gets programmed into one state in the positive cycle, and a different state in the negative cycle. It keeps oscillating between these two stable states, forming the famous eight-shaped pinched hysteresis loop in its V-I characteristics (Refer Fig. \ref{fig_imax} ,\footnote {It is worth noting that the transition in resistance value happens only when the polarity of the current is changed. For other pulses, the resistance remains constant. The device oscillates between two stable resistance-states for the same maximum current in opposite directions. It is evident by looking at regions depicted by A1 to A5 that the change in resistance happens at the first pulse of the transition. One may also note that these observations show recovery of resistance to a higher resistance state: A5 resistance is higher than A4 resistance. These results can be reproduced by careful experimentation for all the three devices.}). It has been established that If it is pinched, it is memristive. Pinched hysteresis loop is the fingerprint of a memristor \cite{kim2012pinched}.  

By using various stimuli with different maximum amplitudes on either sides, the device can be programmed to function in multiple stable resistance-states and move between them. When used as a resistive RAM, the memory can be read in the "memory" state by providing an excitation of a small amplitude. This fulfill the conditions of Chua's definition of memristor, and qualifies the century-old coherer as a canonical implementation of a memristor. 

\section{\label{sec:level1} DISCUSSION} 

We have shown, through new results, that the century old coherer and auto-coherer  function as a multi-state resistance RAM and is thus the canonic implementation of the elusive memristor. It intrigued the science of that era as much as memristor is exciting the scientists of the present day \cite{prodromakis2012two}. The present work shows that one does not require specific material or precise construction to implement memristors. It is a natural property of metallic point contacts. Note that there is another component called memistor which is an entirely different component and must not either be confused with coherer or memristor. It is rather an ill posed 3 terminal device \cite{kim12memistor, xia2011two}

There are certain differences between the  behavior of coherers and other present day memristors. Unlike Williams et. al. memristor \cite{strukov2008missing}, our devices do not behave as a charge-flux based memristor. Irrespective of the increase or decrease of flux, their resistance does not change till the maximum current or current polarity changes. Our device have similarities in behavior \cite{jo2008cmos, kim2010nanoscale} and construction \cite{kim2009nanoparticle} to that of some other memristors recently fabricated at nano-scale. However, none of these recent memristors have reported observation of multiple resistance states or dependence on $I_{max}$.

The question worth discussing is whether the resistive switching mechanism is due to existence of oxide at the interface of the metals. Our preliminary experiments with polished gold balls showed the said behavior, which indicates (but does not rule out) that the observations are not due to presence of oxide. From an analytical standpoint, the construction and behavior of our device doesnÕt fit those observed in oxide based memristors. The construction and mechanism of operation of oxide based memristors is discussed in detail in the review by Waiser \cite{waser2009redox}. One class of oxide based memristors switch unipolarly due to formation and melting of filaments thermally. This is similar to the explanation provided in coherer literature \cite{falcon2010efecto, bquin2010electrical}. Our memristor has bipolar switching and cannot be explained by a thermal process which is independent of current direction. Only the initial cohering action, akin to electroforming step reported in literature, may reasonably be explained by a thermal heating process. \\
Among bipolar oxide-based memristors, one class (Valence Change Mechanism) uses specific transition metal oxides or those with defects, whereas the other class has dissimilar electrodes (one active and one counter electrode) on the two sides of the oxides (or an electrolyte). In the latter case, the difference in the properties of the two electrodes leads to dependence on current direction. Our memristor has no explicitly introduced vacancy defects at the interface, doesnÕt require transition metal oxides and works perfectly well with the same metal across all contacts. Thus, its construction and behavior, put together, do not resemble any oxide based memristor configuration and behavior.\\
On the other hand, ferroelectric RAM containing a perovskite layer and nano-particle assemblies  \cite{kim2009nanoparticle} are symmetric, and yet show bipolar resistance-switching caused by electric field induced polarization. We believe that the behavior of our device is similar and a result of polarization at the contacts formed between the metals. It is still open to investigate whether this happens due to the geometry at the contact or due to impurities. The same requires to be investigated through material analysis and microscopic studies.\\
Our new results show that bipolar switching can be observed in a large class of metals by a simple construction in form of a point-contact or granular media. It does not require complex construction, particular materials or small geometries. The signature of all our devices is an imperfect metal-metal contact and the physical mechanism for the observed behavior needs to be further studied. That the electrical behavior of these simple, naturally-occurring physical constructs can be modeled by a memristor, but not the other three passive elements, is an indication of its fundamental nature. By providing the canonic physical implementation for memristor, the present work not only fills an important gap in the study of switching devices, but also brings them into the realm of immediate practical use and implementation.

\section{\label{sec:level1} ACKNOWLEDGEMENT}

The authors would like to thank Prof. Leon Chua, Prof. Karl Berggren, Prof. Rohit Karnik, Dr. Una-May O'Reilly, Prof. Tamas Roska, Prof. Steve Kang for their help and suggestions. The authors would also like to thank Nimish Girdhar for his help.

\bibliography{ref}

\end{document}